\documentclass[fleqn,10pt]{wlscirep}
\usepackage{subfigure}

\newcommand{\av}[1]{\langle #1 \rangle}

\newcommand{\FigPath}{.}%please keep this command and modify it accordingly.

\title{Emergence of metapopulations and echo chambers in mobile agents}

\author[1]{Michele Starnini}
\author[2]{Mattia Frasca}
\author[3]{Andrea Baronchelli}
\affil[1]{Departament de F\'{\i}sica Fonamental, Universitat de
  Barcelona, Mart\'{\i} i Franqu\`es 1, 08028 Barcelona, Spain}
\affil[2]{Dipartimento di Ingegneria Elettrica Elettronica e Informatica,
University of Catania, Viale A. Doria 6, 95125 Catania, Italy}
\affil[3]{Department of Mathematics, City University London, London EC1V 0HB, UK}

\begin{abstract}
Multi-agent models often describe populations segregated either in the physical space, i.e. subdivided in metapopulations,
or in the ecology of opinions, i.e. partitioned in echo chambers.
Here we show how the interplay between homophily and social influence controls the emergence of both kinds of segregation
in a simple model of mobile agents, endowed with a continuous opinion variable.
In the model, physical proximity determines a progressive convergence of opinions but differing opinions result in agents moving away from each others.
This feedback between mobility and social dynamics determines to the onset of a stable dynamical metapopulation scenario
where physically separated groups of like-minded individuals interact with each other through the exchange of agents.
The further introduction of
confirmation bias in social interactions, defined as the tendency of an individual to favor opinions that match his own,
  leads to the emergence of echo chambers where different opinions can coexist also within the same group.
   We believe that the model may be of interest to researchers investigating the origin of segregation in the offline and online world.
\end{abstract}

\begin{document}

\flushbottom
\maketitle

\noindent Correspondence and requests for materials should be addressed to \\
M.S. (michele.starnini@ub.edu), M.F. (mattia.frasca@dieei.unict.it),  or A.B. (Andrea.Baronchelli.1@city.ac.uk)

\thispagestyle{empty}

\section*{Introduction}

Metapopulation models were first introduced in ecology \cite{ilkka1999metapopulation} and are nowadays widely used for the study of a large variety of phenomena, ranging from epidemic spreading to general social phenomena \cite{watts2005multiscale,colizza2007invasion,liu2013contagion}. They describe agents, possibly of different species, that diffuse stochastically on the vertices of a network and interact among them upon contact on the same vertex, whose occupancy is not limited (hence the term `bosonic' often used to describe this models \cite{baronchelli2008bosonic}). Recently, attention has been devoted to understand how different features of the metapopulation network, fixed during the dynamics of the process, affect the behavior of the system \cite{colizza2007reaction,baronchelli2008bosonic}. However, less attention has been devoted to describe how metapopulation structures \textit{emerge} in absence of obvious environmental factors. How does a population end up into many segregated subgroups?

The pioneering work of Schelling showed that no strong incentives are needed at a microscopic level for segregation to occurr.  Specifically, it showed that individual weak preferences to be surrounded by similar others are sufficient for the emergence of spatially segregated subpopulations, and no strong desire to avoid different others is needed\cite{schelling1969models,schelling1971dynamic}. Further studies confirmed this picture \cite{henry2011emergence}, showing that even a population of individuals who actively seek diversity can result into segregation \cite{zhang2004residential,pancs2007schelling}. All of these models describe individuals that occupy the nodes of a lattice or network and are endowed with a visible trait that rules their interactions (e.g., skin color). They address the question of how different microscopic interaction rules impact the spatial distribution of these individuals. Crucially, the visible trait is modelled as a fixed and immutable characteristic of the individuals.

In this paper, we relax the assumption of a fixed trait of the individuals and focus on the interplay between spatial mobility and opinion dynamics.
We consider mobile agents characterized by a continuous variable, or ``opinion'',
that evolves in time depending on the microscopic interactions between individuals.
The model we propose incorporates

\begin{itemize}
\item Mobility: Individuals move in a two dimensional space;
\item Homophily: Individuals have a tendency to interact more with others who share their opinion;
\item Social Influence: Interactions between individuals tend to increase their similarity.
\end{itemize}

\noindent These three elements have been considered by a number of models, but the interplay between them has received less attention \cite{Castellano09}. For example, the question of why homophily and social influence do not necessarily yield social uniformity motivated the well-known Axelrod model of dissemination of culture \cite{axelrod1997dissemination,castellano2000npt}, and models based on mobile interacting agents have been used to study such phenomena as
epidemic spreading \cite{buscarino2014local},
social cooperation \cite{meloni2009effects}
and consensus \cite{baronchelli2012consensus}.
Here, we consider the mobility scheme introduced in Refs.\cite{PhysRevLett.110.168701, citeulike:13344055},
which reproduces empirical data on human face-to-face interactions in social gatherings.
However, while Ref.\cite{PhysRevLett.110.168701} considers individuals characterized by a quenched ``attractiveness''
ruling the duration of their interactions,
here the role of attractiveness is played by the status or opinion of individuals with respect to the other agents, that is, by a \emph{dynamic} variable.
Individuals tend to become more similar to the agents surrounding them and have a higher probability of moving away from dissimilar peers.
We show that for certain values of the parameters the dynamics of the model yields the \textit{emergence} of a metapopulation structure,
i.e. physically segregated groups of individuals sharing similar opinions, that yet interact with each other through the exchange of agents.

Finally, we address more in detail the relation between the segregation in the physical and opinion spaces. We specify further the role of homophily by introducing a confirmation bias of the agents,
i.e. a tendency to acquire or evaluate new information in a way that confirms one's current opinion and avoids contradiction with prior beliefs \cite{nickerson1998confirmation,allahverdyan2014opinion}.
Following a consolidated line of modelling in the context of continuous-variable models \cite{deffuant2000mixing,hegselmann2002opinion,Castellano09},
we model confirmation bias as a bounded confidence between agents.
This assumption reflects the idea that individuals will not interact with others whose opinion is too dissimilar from their own,
and corresponds to a step-function description of the confirmation bias.
The introduction of bounded confidence leads to the emergence of `echo chambers' of agents
that co-exist within the same group but do not influence each other,
 thus breaking the coupling between spatial segregation and polarization in the space of opinions.

 \section*{Model definition}
\label{sec:model}
The model is defined as follows.
$N$ individuals, initially randomly distributed in a square box of linear size $L$
(corresponding to a density $\rho = N/L^2$), perform a random walk of fixed step length $v\cdot \delta t$ and
interact with the agents they find within a certain distance $d$. For the rest of the paper, without loss of generality, we fix $\delta t=1$.
The position of agent $i$ at time $t$ is indicated as $(x_i(t),y_i(t))$.

Each individual $i$ is characterized by a dynamical state variable $s_i(t) \in [0,1]$
representing his opinion, whose initial value $s_i(0)$ is randomly extracted from a
 uniform and bounded distribution $F(s)$ within the interval $s \in [0,1]$.
 The $s_i$ variable is defined in an opinion space with periodic boundary conditions,
 so that the difference between two states $s_i$ and $s_j$ is always taken by modulo 1.
Upon interaction, individuals modify their status
seeking a local consensus with their neighbors.
At each time $t$ the status of each agent $i$, $s_i(t)$, is updated as

\begin{equation}
s_i(t+1) = s_i(t) + K \sum_{j \in \mathcal{N}_i(t)} \left( s_j(t)-s_i(t) \right ) _{\mathrm{mod} \, 1} \, ,
\label{eq:status_dyn}
\end{equation}

\noindent
where $\mathcal{N}_i(t)$ is the set of neighbors of agent $i$,
i.e., $\mathcal{N}_i(t)=\{ j : (x_i-x_j)^2+(y_i-y_j)^2 = d_{ij}^2 < d^2 \}$, and
$K$ is the coupling constant regulating the strength of the social influence they experience.
However, individuals may also change autonomously their status.
At each time step $t$, each individual has a very small probability $R$ to reset his status $s_i(t)$ to a new value, randomly extracted from the distribution $F(s)$. The reset rate $R$ introduces noise in the model and accounts for external factors or exogenous sources of information that
can influence the opinion of the individuals.

.

Homophily is modeled through the probability that an individual
will continue the interaction with his neighbors or not.
If an individual shares a similar opinion with his neighbors
he will probably remain with them, otherwise he will walk away.
Thus, the walking probability $p_i(t)$ of the agent $i$ at time $t$
is proportional to the difference between his status $s_i(t)$
and the status of his most similar neighbor,

 \begin{equation}
 p_i(t) = \min_{j \in \mathcal{N}_i(t)} | s_i(t) - s_j(t) |_{\mathrm{mod} \, 1} \, \, ,
 \label{eq:motion_rule}
 \end{equation}

\noindent where, as in Eq. \eqref{eq:status_dyn}, the difference between the states is taken modulo 1 to model an opinion space with periodic boundary conditions.
 Thus, each agent maintains his position with probability $p_i(t)$
or performs a step of length $v$ in a random direction with probability $1-p_i(t)$.
Isolated agents have a walking probability $p_i(t) = 1$.
Thus, heterogeneous groups are more fragile as individuals will tend to abandon them.
Conversely, groups whose individuals experience a strong consensus will tend to persist in time with stable status \textit{and} position.

\section*{Results}

The model dynamics is fully characterized by four parameters:
the collision rate $p_c$, given by the product between the density $\rho$
and the interaction area $\pi d^2$, that is $p_c = \pi d^2 \rho$,
the velocity of agents $v$, the coupling constant $K$, and the reset rate $R$.
The results presented here are for numerical simulations of the model
with $N=200$ agents, $L=50$, $K=5 \cdot 10^{-3}$, $R=2 \cdot 10^{-5}$, $v=2$, $d=1$, averaged over $10^2$ runs, unless otherwise specified.
To avoid spurious effects due to a transient state, we start to observe the system after a time $T_0 =10^3$,
which turns out to be sufficient to reach the steady state under different conditions,
and run simulation up to $T_{end} = T_0 + T$, with $T=10^5$ time steps.

\subsection*{Group formation}
\label{sec:groups}

The microscopic rules described above introduce a positive feedback between
mobility and opinion dynamics:
Individuals can reach a local consensus based on proximity in the physical space, Eq. \eqref{eq:status_dyn},
and the achievement of a local consensus favors the persistence of that proximity, Eq. \eqref{eq:motion_rule}. As a consequence, in general
the system reaches a quasi-stationary regime characterized by the presence of metastable groups of individuals.
However, two events can alter the equilibrium of a group and change its composition and/or spatial properties, namely the arrival of a new individual and the spontaneous change of opinion by an agent.
In the first case, either the newcomer's opinion is close to the group's local consensus
and he will settle within the group with high probability, or he will leave,
potentially having weakened the group by causing part of its members to shift their opinions - a scenario that may undermine or destroy the existing group.
The second case is equivalent to the first, with the newly re-set agent that may remain in the group or leave it. This dynamic interplay between the processes of group formation and group fragmentation introduces a rich phenomenology which can be understood in light of two quantities:
the average fraction of moving agents, $\av{N_m}$, and the average size of the groups, $\av{S}$, defined as

 \begin{equation}
\av{N_m} = (TN)^{-1} \sum_{t} N_m(t), \qquad \qquad  \qquad
 \av{S} = T^{-1} \sum_{t} |\mathcal{G}(t)|^{-1} \sum_{i \in \mathcal{G}(t)} N_i(t),
 \end{equation}

\noindent where $N_m(t)$ is the number of isolated and moving agents at time $t$,
$N_i(t)$ is the number of individuals forming group $i$ at time $t$,
and $\mathcal{G}(t)$ is the set of groups at time $t$.

		\begin{figure*}[tb]
  \begin{center}
    \includegraphics[width=0.8\textwidth]{\FigPath/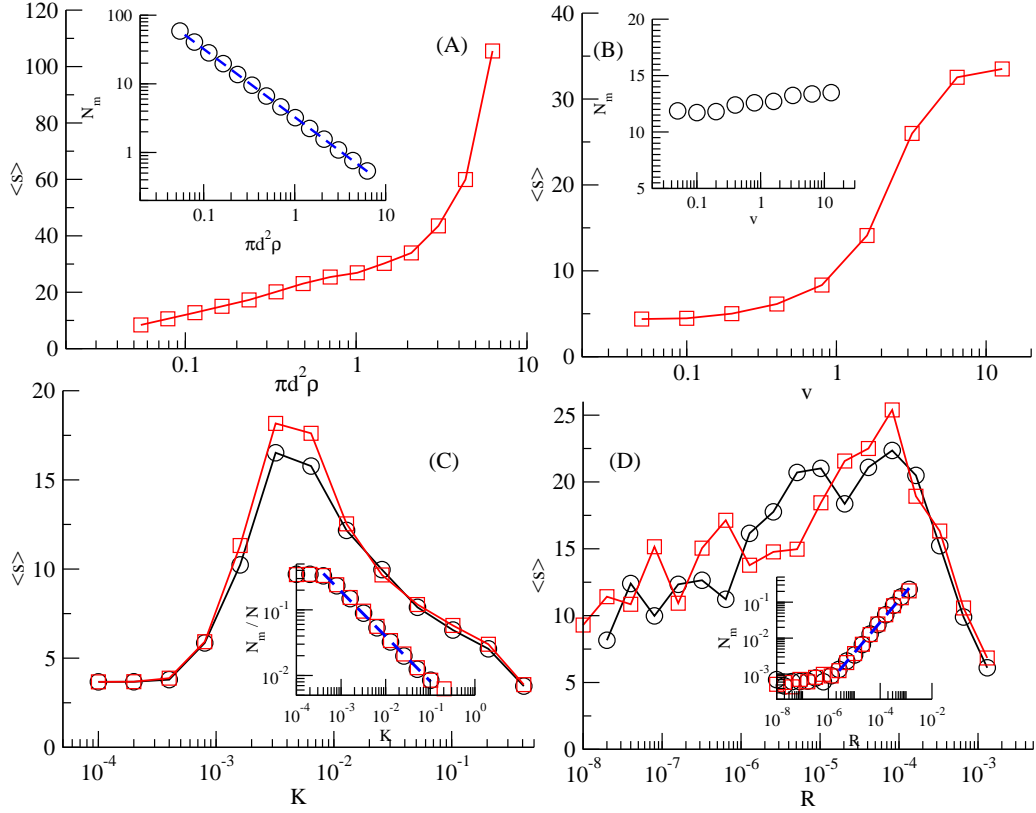}
  \end{center}
  \caption{ \textbf{Behavior of the model as a function of the parameters.}
  Average size of the groups formed $\av{S}$ (main panels) and average fraction of moving agents $\av{N_m}$ (insets).
  (A): $\av{S}$ vs. the collision rate $p_c = \pi d^2 \rho$. 	
  Inset: $\av{N_m}$ vs. $p_c$.
  	The functional form $\av{N_m} \sim p_c^{-1}$ is plotted in dashed line.
  	(B): $\av{S}$ vs. the velocity $v$.	Inset: $\av{N_m}$ vs. $v$.
    	(C): $\av{S}$ vs. the coupling constant $K$.
    	The average size of the groups $\av{S}$ shows a maximum for $K \simeq 0.005$.
    	Inset: Rescaled $\av{N_m}/N$ vs. $K$.
		The functional form $\av{N_m} \sim K^{\alpha}$, with $\alpha = -0.70$, is plotted in dashed line.
    	(D):  $\av{S}$ vs. the  reset rate $R$.	
    	The average size of the groups $\av{S}$ shows a maximum for $R \simeq 10^{-4}$.
    	Inset: Rescaled  $\av{N_m}/N$ vs. $R$.
    	The functional form $\av{N_m} \sim R^{\alpha}$, with $\alpha = 0.85$, is plotted in dashed line.    	    	
    	Panels (C) and (D) show two different sizes of the model, $N=200$ (red squares) and $N=400$ (black circles).  }
  \label{fig:panel}
\end{figure*}

Fig.~\ref{fig:panel} (A) shows the average fraction of moving agents $\av{N_m}$
and the average groups size $\av{S}$
as a function of the collision rate $p_c$.
	For very small collision rates $p_c \ll 1$ the system is formed only by isolated agents and small groups,
	for very large collision rates $p_c \gg 1$ the network formed by the agents percolates,
	and a single group spans a finite fraction of the system,
	while for a large range of intermediate values of the collision rate $p_c  \lesssim 1$
	a regime with several groups of different sizes and few isolated agents moving from one group to another emerges.
	This regime reveals the onset of a metapopulation structure which we further characterize in the next Section.
	The average number of moving agents $\av{N_m}$ is inversely proportional to the collision rate,
 $\av{N_m} \sim p_c^{-1}$ (inset, dashed line).
 This observation is in agreement with the hypothesis that the average number of collisions is constant and it does not depend on the collision rate $p_c$, being regulated only
by the probability that an individual leaves spontaneously a group as determined by $K$ and $R$,
so that $\av{N_m}$ is simply proportional to the expected collision time $\tau_c \sim p_c^{-1}$.

The velocity of agents, $v$, appears to have a small effect on the number of moving agents $\av{N_m}$ % (except for very small $v \ll 1$),
but it is positively correlated with the average group size $\av{S}$,
 large $v$ values leading to significantly larger groups (Fig.~\ref{fig:panel}, B).
 Indeed, large values of $v$ favor the mixing of the system, triggering a sort of rich-get-richer dynamics for the groups where the probability for a new individual to join a group is simply proportional to its size.
 On the contrary, small values of $v$ promote the stability of existent groups,
 as the probability for a walking individual to re-join the group he just left is high.

The average fraction of moving agents $\av{N_m}$
and the average groups size $\av{S}$
show strong dependence on the strength of social influence $K$ (Fig.~\ref{fig:panel}, C).
The larger the value of $K$, the smaller the number of interactions required
before a moving agent reaches a local consensus within some group,
so the number of moving individuals $\av{N_m}$ decreases with $K$
(with a functional form compatible with power law, $\av{N_m(K)} \sim K^{\alpha}$, with $\alpha = -0.71$, see inset).
Note that, even for very small values of $K$, unstable groups of $4-5$ individuals are formed,
and the number of moving individuals $\av{N_m}$ is stable with respect to $K$.
The average size $\av{S}$ slowly increases as $K$ increases, reaches a maximum for $K \simeq 0.005$,
and then it decreases for larger values of the coupling constant $K$.
With large $K$, indeed, the model dynamics produces more pairs, since individuals tend to get stuck with their first encounter.

A similar behavior is observed for the average size of the groups $\av{S}$
as a function of the reset rate $R$.
In this case the peak of $\av{S}$ is smoother, with a large range of values of $R$ for which the size is maximum,
$R \simeq 10^{-6} - 10^{-4}$ (Fig.~\ref{fig:panel}, D).
In fact, low values of $R$ favor the formation of small but very stable groups, while high values of $R$ produce smaller and very unstable groups.
A confirm of this behavior can be found by observing the number of moving agents $\av{N_m}$ as a function of $R$,
shown in the inset.
For small values of $R$ there are few or no moving agents, and the system is almost frozen in a state with small groups of $7-8$ individuals,
while as $R$ increases, the number of moving agents increases as well, with  $\av{N_m(R)} \sim R^{\alpha}$, with $\alpha = 0.71$.

We have also carried out extensive numerical simulations to address the dependence of the model with the number of individuals $N$, concluding that the model behavior is independent of the system size, if the density $\rho$ is kept constant. As an example, in Fig.~\ref{fig:panel} (panels C and D) the curves are plotted for two values of the system size $N=200$ and $N=400$, showing that both the group size $\av{S}$ and the fraction of moving agents $\av{N_m}$ scale with $N$.

\subsection*{Onset of metapopulation structures}
\label{sec:metapopulation}

The metapopulation regime is characterized by groups of individuals heterogeneously distributed in the physical and opinion spaces.
At any time, some individuals leave and join groups, moving freely in the physical space.
This regime emerges when the collision rate $p_c$ is large enough to allow
groups formation and the interchange of individuals between groups,
but smaller than the percolation threshold.
 The size of the groups, as well as the number of individuals moving between groups,
are regulated by the strength of the social influence of the individuals,
modeled by the coupling constant $K$, and the external influence represented by the reset rate $R$
(see Fig.~\ref{fig:panel}, C and D).

\begin{figure*}[t]
  \begin{center}
%\subfigure[]{\includegraphics[width=5.8cm]{\FigPath/trajectA.pdf}}
%\subfigure[]{\includegraphics[width=5.8cm]{\FigPath/trajectB.pdf}}
%\subfigure[]{\includegraphics[width=5.8cm]{\FigPath/trajectC.pdf}}
\includegraphics[width=1\textwidth]{\FigPath/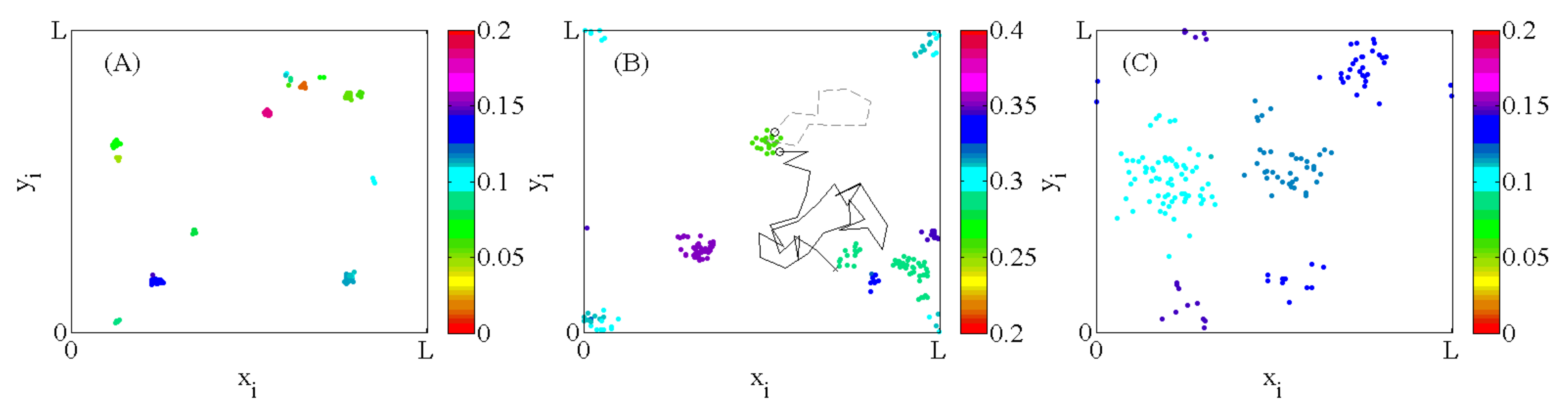}
  \end{center}
  \caption{ \textbf{Sketch of the metapopulation structure emerging in the model.}
  Parameters are set to $N=200$, $R=10^{-4}$, $K=0.01$, $v=2$, and different values of the collision rate:
  (A) $p_c=0.1$ (for a full video, see Movie S1), (B) $p_c=1$ (Movie S2), (C) $p_c=5$ (Movie S3). A time window of 1500 steps is considered. Agents not moving in this time window (i.e., forming stable groups) are represented as filled circles whose color codes for the status value. In addition, for panel B, the trajectories of two agents moving in the time window considered are also drawn. One of them (continuous line) moves from one group into another, while the other (dashed line) comes back to his own starting group.
  \label{fig:metapopulation}}
\end{figure*}

The metapopulation regime is illustrated in Fig.~\ref{fig:metapopulation}, showing a snapshot of the model behavior
for three different values of $p_c$, modified by varying the density of agents (see also Movies S1, S2 and S3). Agents
 not moving in a time window of $1,500$ timesteps, i.e., forming stable groups over that observation time, are represented as filled circles.
 The value of agent status is color coded,
illustrating how groups are formed by individuals sharing similar opinions.
We note that as $p_c$ increases, the number of groups reduces and their size increases.
At the same time, the difference in the opinions between the groups reduces,
indicating that the system is approaching the regime where a unique group of individuals sharing the same opinion is formed.
Panel B also shows the trajectories of two agents selected among those who have changed position at least once in the time observation window. They are representative of two distinct cases. One of them initially belonged to a group located in the right, bottom part of the physical space, then left the group and joined another one approximatively located in the center of the figure (trajectory depicted as a continuous line). The second individual left the group located in the center of the plane but eventually joined it again to remain within it for the rest of the observation (dashed line).

In literature, metapopulation structures are characterized by the connectivity matrix representing metapopulations as nodes of a graph,
 and defining which groups may exchange agents.
 In our system, the connectivity is the result of the dynamical self-organization of the system.
 Despite the fact that every link is possible, only some of them, depending on the system parameters, are statistically relevant.
 In particular, the typical scenario observed for low values of $v/L$ is that agents leaving a group travel for a short distance before reaching the new group.
 The probability that this group is one of the closest (in a geographical sense) is high,
 yielding a group connectivity that can be modeled as a random geometric graph.
 On the other extreme, for high values of $v/L$, agents are allowed to reach potentially any other group starting from their own,
thus we expect a connectivity pattern with an all-to-all coupling between the populations.

The characteristic timescale over which a metapopulation does not change its state depends on
the reset rate $R$ and on the coupling constant $K$.
A large reset rate $R$ implies a shorter lifetime for a group, while
a large coupling strength $K$ increases the resilience of a group to perturbations
such as the spontaneous change of opinion by a member or the arrival of a new individual.
Also the collision probability $p_c$ impacts the stability of groups:
for small values of $p_c$ few individuals will collide with formed groups,
intermediate values of $p_c$ will produce more collisions and less stability,
and for very large values of $p_c$ the system percolates into one or few big groups
which tend to keep their composition unchanged in time.

\begin{figure}[tbp]
\begin{center}
\includegraphics*[width=8cm]{\FigPath/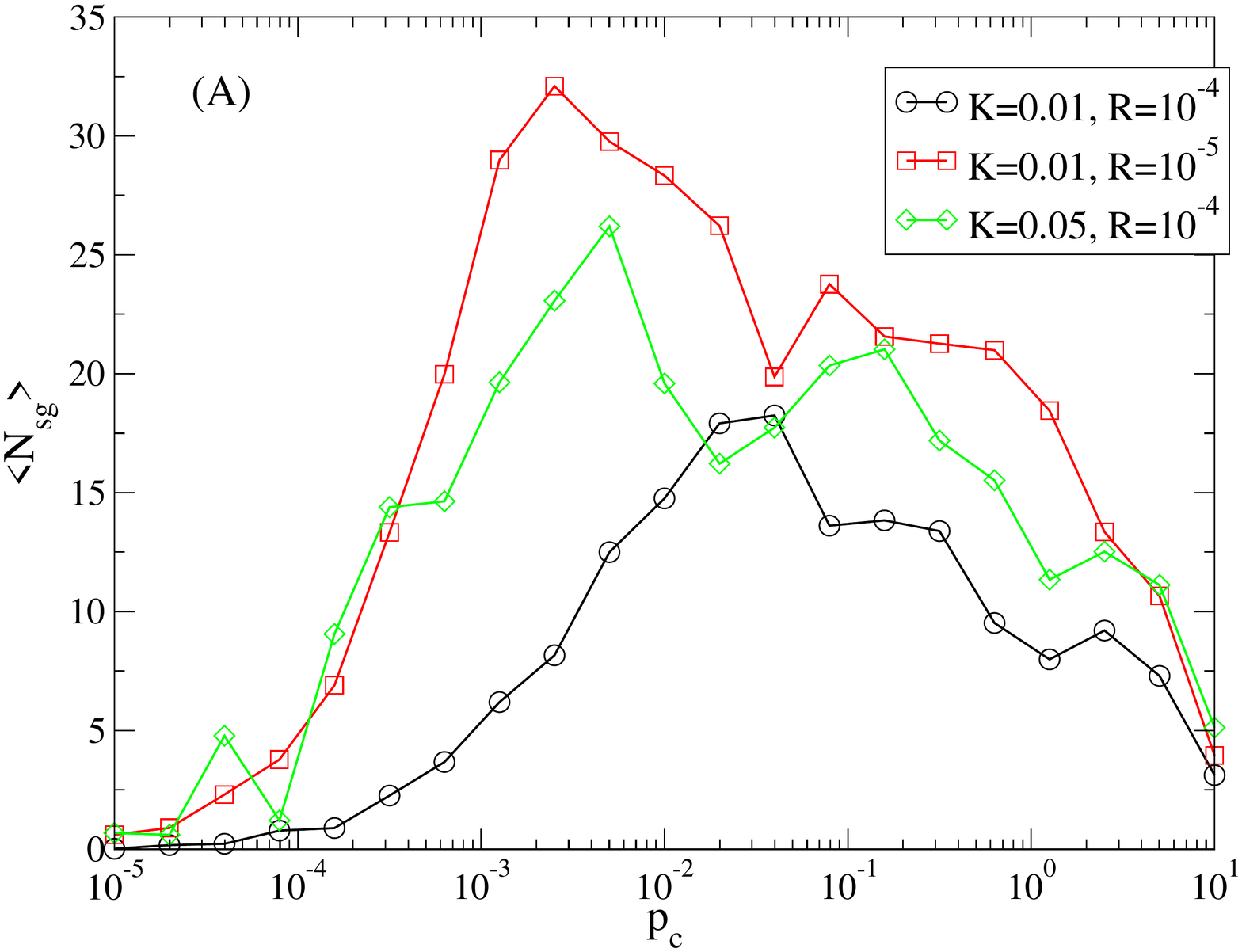}
\includegraphics*[width=8cm]{\FigPath/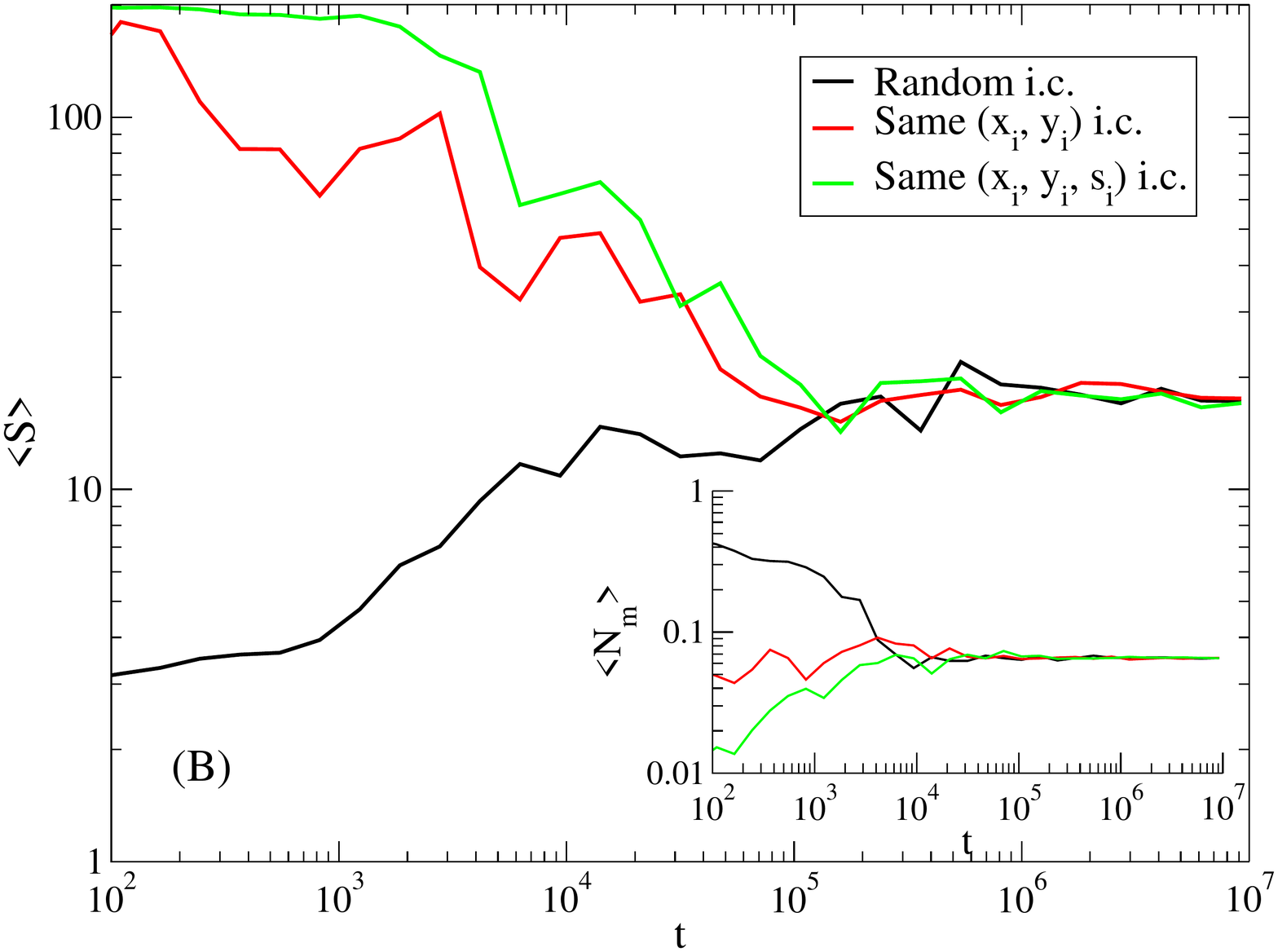}
\end{center}
\caption{ \textbf{Stability of groups and effect of initial conditions.}
(A): Number of stable groups $\av{N_{sg}}$, calculated in a time window of length $\tau=50$, as a function of the collision rate $p_c$,
for different values of $K$ and $R$.
(B): Average groups size $\av{S(t)}$ (main)
   and average number of moving individuals $\av{N_m (t)}$ (inset) as a a function of time $t$,
   for different initial conditions: random values of position and status,
   same position and random status, same position and status.
\label{fig:stablegroups}}
\end{figure}

This behavior is illustrated in Fig.~\ref{fig:stablegroups} (A), showing the number of stable groups in a given time window of length $\tau$, indicated as $\av{N_{sg}}$\footnote{The number $\av{N_{sg}}$ of groups that are stable in a given time window of length $\tau$ is calculated as follows. At each time instant, we build the adjacency matrix $\{A_{ij}(t)\}$ with $A_{ij}(t)=1$ if $i$ and $j$ are close in the physical space, and $A_{ij}(t)=0$, otherwise. From $\{A_{ij}(t)\}$, for $t>T_0+\tau$ we construct the matrix $\{\mathcal{A}_{ij}(t)\}$, which now indicates if two agents have been neighbors in the whole time window $[t-\tau+1,t]$, that is, $\mathcal{A}_{ij}(t)=1$ if ${A}_{ij}(h)=1$ for $h=t-\tau+1,\ldots,t$, and $\mathcal{A}_{ij}(t)=0$, otherwise. We then calculate $N_{sg}(t)$ as the number of components of $\{ \mathcal{A}_{ij}(t)\}$ and average over time to obtain $\av{N_{sg}}$.}, for selected values of the parameters ruling the model.
As expected,  $\av{N_{sg}}$ grows with $p_c$ until it reaches a maximum and
then decreases as the system percolates into one or few groups.
Therefore, the metapopulation regimes emerges for collision rates $p_c$ between
the extreme scenarios of very small groups and system percolation.
Fig.~\ref{fig:stablegroups} (A) also shows how the average number of stable groups is affected by
the coupling strength $K$ and the reset rate $R$.
Increasing $K$ has an effect qualitatively similar to that obtained by a decrease of $R$
(compare the curve for $K=0.05$, $R=10^{-4}$ with that for $K=0.01$, $R=10^{-5}$) as, in both cases,
the formation of groups with few individuals is favored and
a large number of stable groups appears even at low values of $p_c$.
Therefore, the strength of the social influence $K$ and that of the external opinion sources $R$
act as opposite forces on the group stability.

Our analysis also focused on the role of initial conditions.
Does a single stable group segregate into several different subgroups under the dynamics driven by a nonzero reset rate $R$?
Or rather do particularly homogeneous initial conditions guarantee asymptotic stability?
To address this question we consider two different kinds of extreme initial conditions: (i) all the agents have the same position in the space,
$x_i = y_i = L/2$, $\forall i$, and (ii) all the agents have the same position in the space and the same status value,
$x_i = y_i = L/2$, $s_i(0) = 0.5$, $\forall i$.
Interestingly, the feedback between the dynamics in the physical space
and the space of social consensus is enough to partition the population into different groups, well separated in both physical and opinion space, under both initial conditions, reaching a dynamical equilibrium indistinguishable from the one achieved with random initial conditions.
Fig.~\ref{fig:stablegroups} (B) shows the evolution in time of the average number of moving individuals,
$\av{N_m (t)}$, and the average groups size, $\av{S(t)}$, for random initial conditions, compared to initial conditions i) and ii).
While initially the evolution of these quantities is different, in the large time limit, $T \gtrsim 10^6$, both $\av{N_m (t)}$ and $\av{s(t)}$ converge to the values that do not depend on the initial conditions. The model is therefore robust to changes in the initial conditions of the population.

\subsection*{Emergence of echo chambers}
\label{sec:echochambers}

In the previous Sections, homophily was implemented through the probability of motion  $p_i(t)$
defined so as to favor repeated interactions between agents sharing similar opinions.
In this Section, we enrich the picture and consider confirmation bias,
i.e., the tendency to prefer and select information in a way that confirms one's beliefs or hypotheses,
while giving less consideration to alternative possibilities  \cite{nickerson1998confirmation}.
We operationalize the confirmation bias through a parameter of bounded confidence,
according to which individuals are influenced only by peers whose opinion is not too different
 from their own \cite{deffuant2000mixing,hegselmann2002opinion,Castellano09}.
 Thus, within this framework, homophily enters also in the definition of the interaction rule between individuals \eqref{eq:status_dyn},
  as social influence between agents $i$ and $j$ now depends on the difference of opinions:

\begin{eqnarray}
s_i(t+1) &=& s_i(t) +  \sum_{j \in \mathcal{N}_i(t)} K_{ij} \left( s_j(t)-s_i(t) \right)_{\mathrm{mod} \, 1} ,  \\
\mbox{with} \; \; \; K_{ij} &=& \begin{cases} K & \mbox{if }  |s_j(t)-s_i(t)|_{\mathrm{mod} \, 1} \leq \frac{1}{2}(1-C) \nonumber  \\
0 & \mbox{ otherwise.} \end{cases}
\label{eq:status_dynmodifiedforCB}
\end{eqnarray}

\begin{figure}[tbp]
  \begin{center}
\includegraphics[width=8cm]{\FigPath/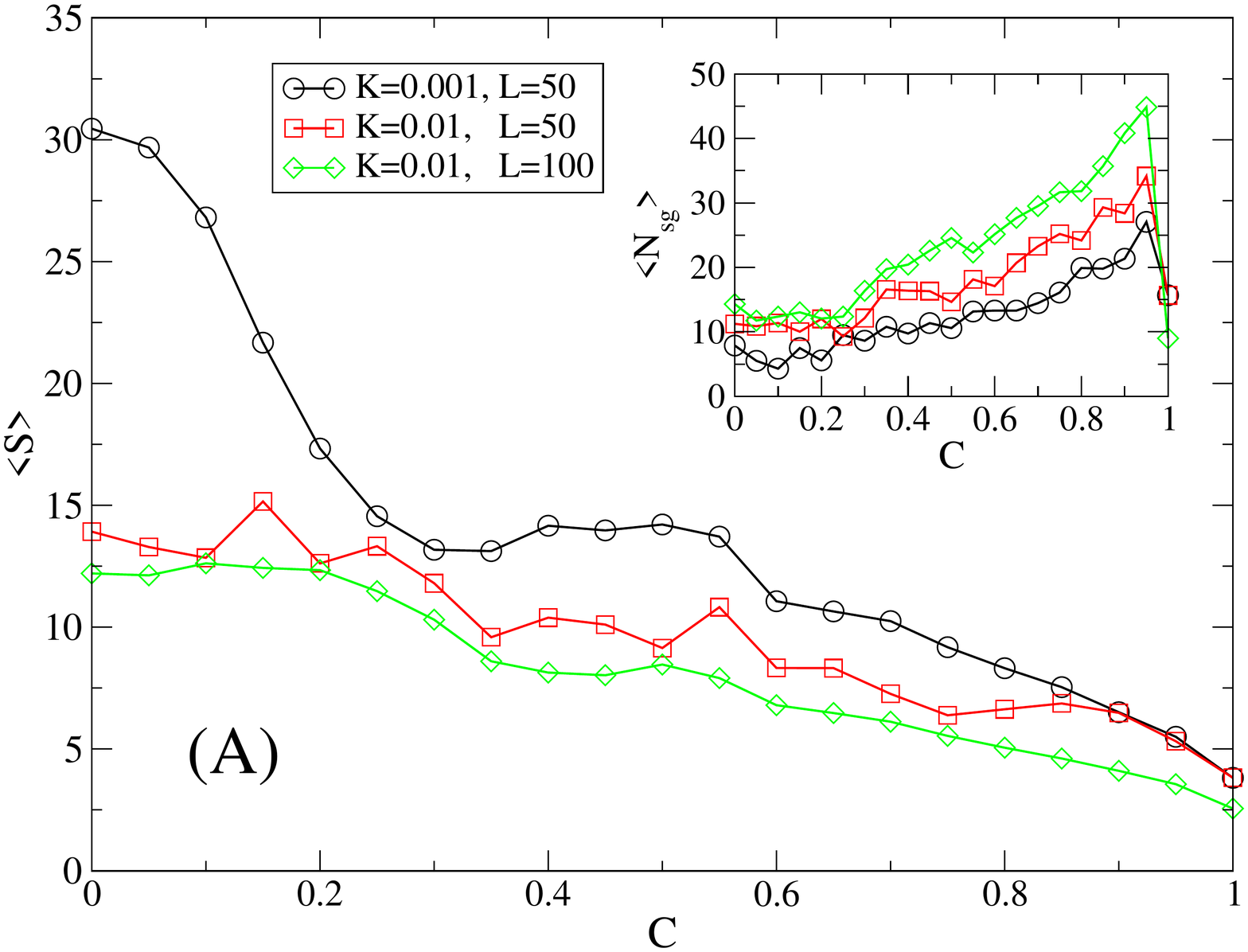}
\includegraphics[width=8cm]{\FigPath/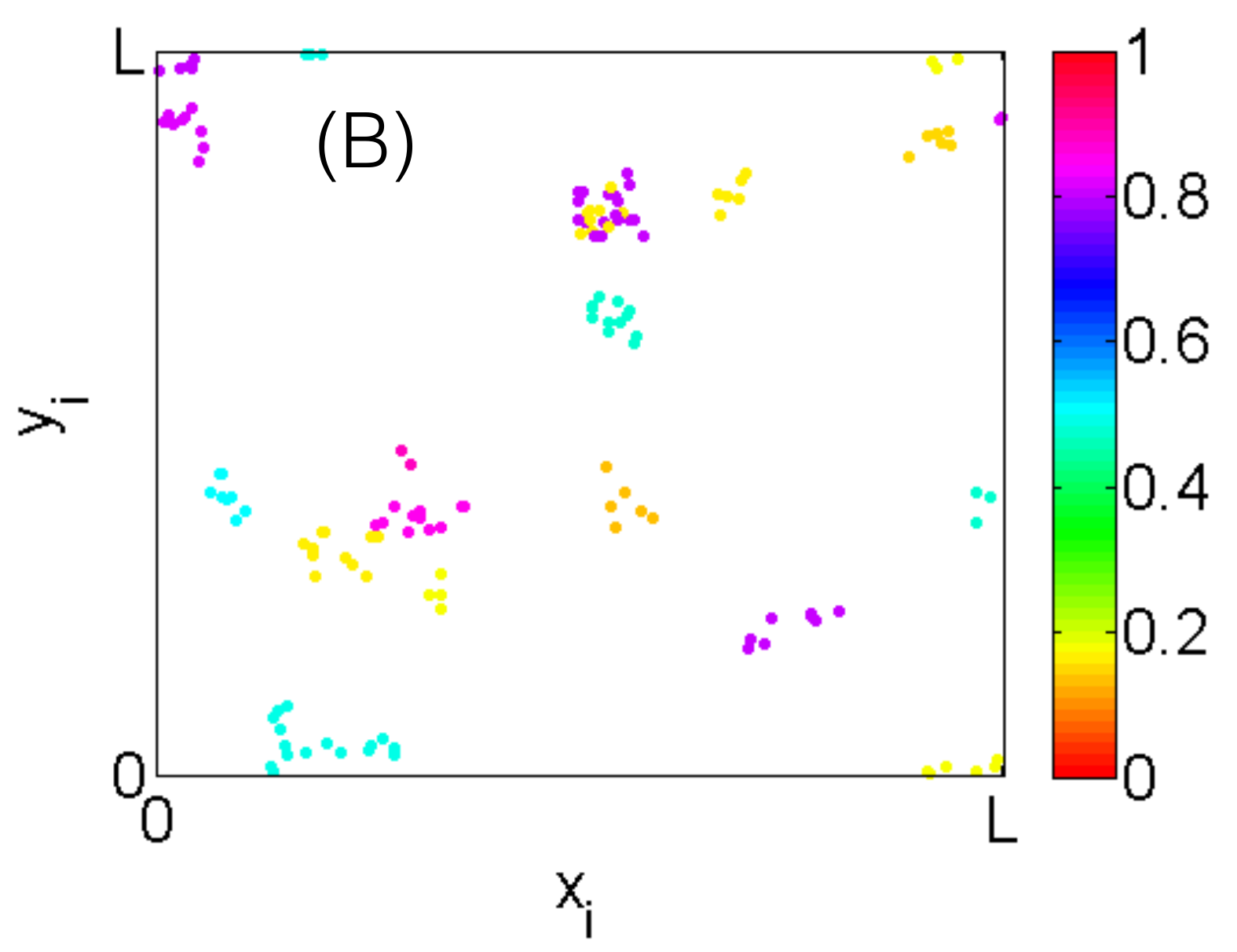}
%\subfigure[]{\includegraphics[width=0.6\columnwidth]{\FigPath/traiettorieCB.pdf}}
  \end{center}
  \caption{ \textbf{Effects of confirmation bias.}
 (A): Average group size $\av{S}$ (main), and number of stable groups $\av{N_{sg}}$ (inset),
  as a function of the parameter $C$ tuning the strength of the bias,
  for different values of the social influence strength $K$ and collision probability $p_c = \pi d^2 \rho$.
  tuned by varying the box size $L$.
  (B): Sketch of the metapopulation structure obtained for $C=0.6$ (for a full video, see Movie S4).
    The other parameters are fixed as in Fig.~\ref{fig:metapopulation} (b). 
    Note the presence of groups formed by agents with very different opinions.  
        \label{fig:confirmationbiasSandNsg}}
\end{figure}

%\end{equation}

\noindent Here, the parameter $C \in [0,1]$ tunes the strength of the confirmation bias.
For $C=0$ we recover Eq. \eqref{eq:status_dyn} (note that, giving periodic boundary conditions, the status difference cannot be greater that 0.5),
while increasing values of $C$ represent a stronger bias, the larger $C$ the more an individual is biased toward similar opinions,
so that fewer individuals are able to influence him.
In the limit $C \rightarrow 1$  social influence vanishes since no agent is able to influence any other one.

%Note that Eq. \eqref{eq:motion_rule}

 Fig.~\ref{fig:confirmationbiasSandNsg} (A) shows the average group size, $\av{S}$, 
and the number of stable groups, $\av{N_{sg}}$, as a function of $C$,
showing that a stronger confirmation bias yields a larger number of stable groups, yet their size $\av{S}$ decrease with $C$.
This behavior is common across different values of the parameters, such as the collision rate $p_c$ or the strength of the social influence $K$.
For $C \rightarrow 1$, the number of stable groups drops sharply, due to disappearance of the social influence effect. 
 More importantly, Fig.~\ref{fig:confirmationbiasSandNsg} (B) (see also Movie S4) shows that confirmation bias leads to the presence of groups
 where agents with heterogeneous opinions coexist, i.e. echo chambers of agents that ignore each other although being in spatial proximity
 and belonging to the same (spatially defined) group.

\begin{figure}[tbp]
  \begin{center}
    \includegraphics*[width=8cm,angle=0]{\FigPath/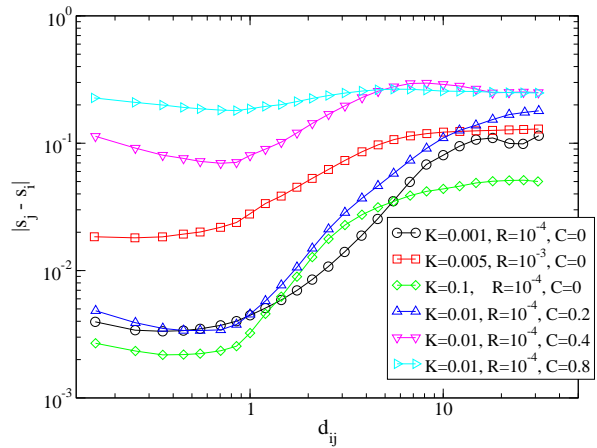}
  \end{center}
  \caption{\textbf{Relation between physical and opinion spaces.}
  Average correlation between the Euclidean distance between two individuals $i$ and $j$, $d_{ij}$,
     and the difference between their opinions, $|s_i-s_j|_{\mathrm{mod} 1} $. Different choices of the parameters of the model are shown. 
     \label{fig:dist_cons} }
\end{figure}

Finally, we address more in detail the close relation between the physical and opinion spaces in order to uncover the impact of the confirmation bias.
In the original formulation of the model, the physical metapopulation structure is mirrored in the space of consensus
as different groups in general experience a consensus on different opinions
(see Fig.~\ref{fig:metapopulation}).
The introduction of confirmation bias changes this scenario, allowing the presence of groups with individuals
sharing different opinions (see Fig.~\ref{fig:confirmationbiasSandNsg} (B)).
This fact can be quantified by measuring the correlation between the Euclidean distance between two individuals $i$ and $j$, $d_{ij}$,
and the difference between their status values, $|s_i-s_j|_{\mathrm{mod} 1} $, shown in Fig.~\ref{fig:dist_cons}.
Without confirmation bias (for $C=0$), for any choice of the other parameters
the difference $|s_i-s_j|_{\mathrm{mod} 1} $ is small and constant for $d_{ij} \leq 1$, i.e. within the radius of interaction.
Then $|s_i-s_j|_{\mathrm{mod} 1} $ grows with the distance and reaches a second plateau for large distances $d_{ij} \lesssim L$,
indicating that only individuals within the same group share the same opinion.
The choice of the parameters impacts the level of consensus:
larger values of the reset rate $R$ or smaller values of the strength coupling $K$
yield less consensus among the agents, and viceversa. The presence of confirmation bias lowers the degree of consensus within groups,
leading to larger values of the difference  $|s_i-s_j|_{\mathrm{mod} 1} $ for small distances $d_{ij}$.
This is due to the fact that two or more echo chambers are formed within the same group, 
formed by agents with different opinions unable to influence each others.
Increasing values of $C$ make the difference $|s_i-s_j|_{\mathrm{mod} 1} $ more insensitive with respect to the distance $d_{ij}$,
and in the limit $C \rightarrow 1$ the opinion difference is independent of the physical distance.

\section*{Conclusions}
\label{sec:conclusions}

This paper studied the interplay between mobility, homophily and social influence
in the context of a simple model of opinion dynamics.
The combination of these ingredients leads to the emergence of
a metastable metapopulation structure in which
groups of like-minded individuals spontaneously segregate in space,
while single individuals constantly leave or join them.
The emergence of the metapopulation regime crucially depends on the density of the agents,
but it is not influenced by the initial conditions.
The metapopulation structure is controlled, in terms of group sizes and stability, by the strength of the social influence, $K$, and
the reset rate, $R$, at which the individuals spontaneously change their opinion.
The feedback loop between mobility and social influence yields a strong assortativity between physical and opinion space:
the closer two individuals are in space, the closer will be their opinions.
This scenario is transformed by the introduction of confirmation bias. The fact that individuals can be influenced only by peers sharing similar opinions leads to the emergence of echo chambers where polarized opinions coexist within the same group.

The contributions of the model are threefold.
First, it shows that spatial segregation can result from
a dynamics involving agents seeking consensus on a non-quenched variable.
Second, it provides a framework in which the metapopulation structure,
often assumed in the modelling of social systems,
emerges from the microscopic rules of the model itself. Third, it shows that confirmation bias yields the possibility that different opinions coexist within the same metapopulation.

We believe that our work opens interesting possibilities of future research.
For example, it would be interesting to investigate the features leading to
the observed fractal patterns of human space occupancy in
archeological records \cite{d2012human,d2013identifying} or
present-day distribution of cities \cite{arcaute2015constructing,arcaute2015hierarchical}.
At the same time, the emergence of metapopulation structures in which
like-minded individuals are relatively isolated from the rest of the subpopulation is interesting also in light of
the recently documented emergence of online echo chambers,
in which misinformation spreads and persists \cite{bessi2015science,del2016spreading,garrett2009echo}.

\bibliography{Bibliography}

\section*{Acknowledgements}

M.S. acknowledges financial support from the James S. McDonnell Foundation. M.F. acknowledges the partial support from the FIR 2015-2016 project of the University of Catania.

\section*{Author contributions statement}
M.S., M.F. and A.B. conceived the research,  M.S. and M.F. performed the simulations, M.S., M.F. and A.B. analyzed the data and wrote the manuscript.

\section*{Additional information}

\textbf{Competing financial interests:} The authors declare no competing financial interests.

\end{document}